\documentstyle[prb,aps,multicol,epsf]{revtex}

\newcommand{\startlongequation}{
\end{multicols}\vspace*{-3.5ex}{\tiny\noindent
\begin{tabular}[t]{c|} \parbox{0.493\hsize}{~} \\ \hline \end{tabular}} }
 
\newcommand{\stoplongequation}{
{\tiny\hspace*{\fill}
\begin{tabular}[t]{|c}\hline\parbox{0.49\hsize}{~} \\ \end{tabular}}
\vspace*{-2.5ex}\begin{multicols}{2}}

\begin{document}
\widetext
\draft

\title{Theory and applications 
of the stress density}
\author{Alessio Filippetti$^{1,2}$ and Vincenzo Fiorentini$^{1,3}$}
\address{{\it (1)}\ Istituto Nazionale per la Fisica della Materia 
and Dipartimento di 
 Fisica, Universit\`a di Cagliari,  Italy\\
{\it (2)}\ Department of Physics, University of California,
Davis, CA 95616, U.S.A.\\
{\it (3)}\ Walter Schottky Institut, Technische Universit\"a{}t M\"u{}nchen,
Am Coulombwall, D-85748 Garching bei M\"u{}nchen, 
Germany}
\date{\today}
\maketitle

\begin{abstract}
Drawing on the theory of quantum mechanical stress,
we introduce the stress density in density functional theory,
and give specific prescriptions for its
practical and efficient implementation in the plane wave
ultrasoft pseudopotential method within the local-density
approximation.
In analogy with  the Chetty-Martin
energy density, the stress density provides a 
spatial
resolution of the contributions to the integrated macroscopic
stress tensor. While this resolution is inherently non-unique
(gauge-dependent), there exist gauge-{\it independent}
ways of using it in practice. Here we adopt the following ones: 
{\it a)} calculating integrated macroscopic stresses over
appropriately defined parts of  a system;
{\it b)} analyzing macroscopic averages of the stress density;
{\it c)} analyzing {\it changes} in the stress density
in response to external perturbation. The abilities of the stress density
are demonstrated for a set of representative test cases from surface
and interface physics: in perspective, the stress density emerges 
as vastly more powerful and predictive than the integrated 
macroscopic stress.
\end{abstract}
\begin{multicols}{2}
\section{INTRODUCTION}\label{intro}
\noindent

Energy and stress are quantities of obvious fundamental importance to
the physics of solid state systems. 
In the last fifteen to twenty years, it has been possible to calculate 
them with progressively increasing accuracy from first-principles using
density-functional theory (DFT).\cite{gross}
The energy and stress as output from a typical calculation are
a scalar and, respectively, a tensor whose values pertain
to an entire finite system or to the unit cell of a periodic system. 
The relevance of the information they carry is concealed in their 
parametric dependence on external constraints, such as atomic geometries. 
It is not difficult, on the other hand, to picture situations in which 
access to local information related to energy or stress would be highly 
desirable. Clearly, such access would require a procedure allowing to 
sort out contributions to energy or stress due to specific regions inside 
the simulation cell.

By way of example, consider the calculation of the energy or stress 
change due to surface formation by means of the common repeated-slab
supercell technique. The energy (or stress) of the slab contain
several informations coupled to one another, which one would like to
access separately. A simple case is that of a symmetric slab: 
to extract the properly surface-related quantities, the bulk contributions 
must be subtracted out, and this requires an independent bulk calculation.
A qualitatively more  complex case is that of an inherently
asymmetric slab, such as frequently found in polar semiconductor
surface simulations: in that case, two inequivalent surfaces are
present, and their  individual surface energies {\it  cannot}
be calculated independently. [It is occasionally possible to symmetrize
the simulation cells in such a way as to obtain the surface energy of
one of the surfaces, but  even then only against a
major  computing effort.]  Clearly, more basic issues may be addressed 
by a local energy or stress analysis: if a surface under stress 
reconstructs, and stress is the alleged driving force, it would be 
enlightening to know which parts of the systems tend to reduce rather 
than increase their stress during the transformation. 

Chetty and Martin,\cite{cm} looking for conceptual as well as
practical solution to these  and similar issues, introduced the
energy density in the DFT framework. The energy density is  
a position-dependent density-functional scalar field whose integral
upon the unit cell equals by construction the total energy.  
In general, this quantity is gauge dependent, {\it i.e.} not uniquely 
defined. Indeed, due to the long-range Coulomb interactions, the
correspondence between a region of space and the energy stored
therein is arbitrary, unless the value of  the functional is fixed 
by symmetry at the boundaries of the region. 

In an isolated system, the energy density is zero at 
the boundaries and outside, and the integral over the region enclosed 
by the boundaries is unique, and equals by construction the total energy.  
A more interesting situation where these requirements are
satisfied is a surface (or interface) slab, infinite and
periodic in the surface plane, and finite in the orthogonal
direction. If its thickness is sufficient that bulk-like
behavior is recovered in its interior, the energy associated with
regions bounded by bulk layers is uniquely fixed and equal to the
bulk energy (also relevant to the simple operative definitions\cite{fm} 
of surface energy). The vacuum contribution is zero by definition. 
Thus, the energy density in the surface regions obeys the appropriate 
boundary conditions, hence its integral upon that region is uniquely 
determined and gauge invariant. Similar arguments apply to interface slabs, 
that couple two different bulk regions and one (or quite possibly two, 
independent) interface region(s).

The concept of quantum-mechanical microscopic stress was introduced in 
the early days of quantum field theory,\cite{old} and revisited and 
deeply investigated more recently.\cite{nm,nm1,nm2,fol}
In this paper, building upon the modern microscopic quantum-mechanical 
theory of stress\cite{nm1,nm2,fol} and drawing on the Chetty-Martin 
theory of the energy density,\cite{cm} we construct a position-dependent 
stress density. We start by defining the stress density as a rank-2 
cartesian tensor field ${\cal T}_{\alpha\beta}({\bf r})$ whose integral 
upon the cell equals by construction the macroscopic stress:
\begin{equation}
T_{\alpha\beta} = \int d {\bf r}\>{\cal T}_{\alpha\beta}({\bf r})\>.
\label{prima}
\end{equation}
The formulation we present below is adapted to the practical 
plane-wave pseudopotential computational framework.  
The alternative definition of a stress field related to the force 
field $F_{\beta}({\bf r})$ through
\begin{equation}
F_{\beta}({\bf r})=-\nabla_{\alpha}{\cal S}_{\alpha\beta}({\bf r}),
\label{field}
\end{equation}
\noindent
cannot be used in this framework, because it fixes
 ${\cal S}$ only up to a gauge (the curl of an arbitrary function). 
While the gauge choice is unrestricted in principle, 
only the Maxwell gauge\cite{nm1}  
is practically tractable for the stress field: 
that gauge  satisfies 
Eq. (\ref{field}) only for systems with purely coulombic interactions,
 conflicting with the fact that pseudopotentials
 produce non-coulombic interactions. 

Numerous applications can be envisaged for the stress density. 
Those we focus on here concern surface and interface physics,
as indeed  structural instabilities and specific electronic effects
at surfaces are fairly commonly associated with (among others) 
the surface excess stress.\cite{ibach}
Three point are worth some attention. First, a spatially-resolved 
stress function can be used to disentangle the macroscopic, integrated 
stress of inequivalent surfaces (or interfaces) that may be coupled 
into a single simulation cell, as is frequently the case for semiconductor 
polar orientations. Second, residual bulk contributions to the 
integrated stress can be computed from the stress density in the same 
calculation in a technically consistent fashion.
Third, but definitely not least, the stress density can be applied as 
a ``stress microscope'' to investigate the role of stress in
specific surface or interface processes; this kind of analysis, 
not relying merely on integrated macroscopic stress, can change 
{\it qualitatively} our understanding of the relation of surface stress 
with ({\it e.g.}) reconstruction\cite{me1,me2} and adsorption.

The plan of the work is as follows: in Sec. \ref{sdf} we formulate 
the stress density for density-functional theory in the local density
approximation (LDA), for the specific instance of the plane-wave 
pseudopotential framework. In Sec. \ref{applica} we test the
functionalities and accuracy of the stress density:
specifically, in Sec. \ref{nps} we apply it to the Al (111) surface, 
in Sec. \ref{ps} to the the polar (100) and (111) surfaces of GaAs, 
and in Sec. \ref{inter} to  the (001) and (110) Ge/GaAs interfaces.
Finally in Sec. \ref{micro} we discuss an application of the stress
density as a microscopic analysis tool in the case of metal surface 
reconstructions. Atomic units are used throughout.  

\section{STRESS DENSITY  IN DFT-LDA}
\label{sdf}

\subsection{Detailed formulation}

Nielsen and Martin \cite{nm1} carried out the quantum formulation of 
macroscopic stress for a general system of interacting particles. 
They also presented \cite{nm2} an expression suitable for the DFT-LDA 
framework, and the explicit form of the  stress tensor  for a
plane-wave basis and  pseudopotentials (a generalization exists 
\cite{dalcor} to gradient-corrected  exchange-correlation functionals).
An analogous formulation for an LCAO basis  was provided by
Feibelman.\cite{feib} 

The Nielsen-Martin expression for the macroscopic stress is comprised
of kinetic, exchange-correlation,  electron-electron interaction, and
electron-ion interaction term. For each of these contributions, we
derive a space-resolved expression whose integral over the simulation 
cell equals the corresponding macroscopic quantity. 

The macroscopic kinetic stress can be  expressed in
two equivalent ways, which reflect the gauge dependence of 
the kinetic stress density. The symmetric form is
\begin{eqnarray}
T^{\rm kin}_{\alpha\beta}& =&
e^2\sum_{\nu{\bf k}}f_{\nu{\bf k}}\omega_{\bf k}\int d{\bf r}\,
\nabla_{\alpha}\psi^*_{\nu{\bf k}}({\bf r})\nabla_{\beta}
\psi_{\nu{\bf k}}({\bf r}),
\label{kin}
\end{eqnarray}
and the antisymmetric form is
\begin{eqnarray}
T^{\rm kin}_{\alpha\beta}& =&-e^2\sum_{\nu{\bf k}}f_{\nu{\bf
k}}\omega_{\bf k} \int d{\bf r}\,\psi^*_{\nu{\bf k}}
({\bf r})\nabla_{\alpha}\nabla_{\beta}\psi_{\nu{\bf k}}({\bf r}),
\label{kin0}
\end{eqnarray}
where $f_{\nu{\bf k}}$ and $\omega_{\bf k}$ are occupation numbers and
k-point weights, respectively.  The symmetric  expression in Eq.
(\ref{kin}) is preferable computation-wise as it does not involve
second derivatives of the wavefunctions (for other reasons to
prefer this form, see Ref. \cite{cm}). Thus we define
the kinetic stress density as
\begin{eqnarray}
{\cal T}^{\rm kin}_{\alpha\beta}({\bf r})={e^2\over 2}\sum_{\nu{\bf k}}
f_{\nu{\bf k}}\omega_{\bf k}[\nabla_{\alpha}\psi^*_{\nu{\bf k}}({\bf r})
\nabla_{\beta}\psi_{\nu{\bf k}}({\bf r})+c.c.].
\label{kin1}
\end{eqnarray}
The calculation of ${\cal T}^{\rm kin}_{\alpha\beta}({\bf r})$ is 
computationally inexpensive. 
The wavefunction derivatives are promptly calculated in Fourier space 
and then carried back to real space by fast-Fourier transform (FFT).

The exchange-correlation contribution to the macroscopic stress is
\begin{equation} 
T^{\rm xc}_{\alpha\beta}=-\delta_{\alpha\beta}\int d{\bf r}\> 
\rho({\bf r}) [\epsilon_{xc}({\bf r})-V_{xc}({\bf r})],
\label{pippo3}
\end{equation}
\noindent
where $\epsilon_{xc}$ and $V_{xc}$ are the exchange-correlation energy
density and the exchange-correlation potential, respectively,
and $\rho$ is the electronic charge density. 
The corresponding expression for the stress density follows
straightforwardly: 
\begin{equation}
{\cal T}^{\rm xc}_{\alpha\beta}({\bf r})=-\rho({\bf r})[V_{xc}({\bf r})-
\epsilon_{xc}({\bf r})]\delta_{\alpha\beta}.
\label{xc}
\end{equation}

The macroscopic Hartree stress is given by
\begin{eqnarray}
T^{\rm e-e}_{\alpha\beta}={e^2\over 2}\int\int d{\bf r}\>d{\bf r}'\rho({\bf r})
\rho({\bf r}'){\partial 
\over \partial_{\alpha\beta}}{1\over |{\bf r}-{\bf r}'|}
\nonumber
\end{eqnarray}
\begin{equation}
=-{e^2\over 2}\int\int d{\bf r}\>d{\bf r}'\rho({\bf r})
\rho({\bf r}'){(r-r')_{\alpha}(r-r')_{\beta}\over |{\bf r}-{\bf r}'|^3}.
\end{equation}
Due to the long-rangedness of coulombic interactions, there is no
unique way to fix the stress density as a functional of the charge density. 
However, an  optimal choice is the Maxwell gauge,\cite{cm,nm1} which
 is physically transparent and computationally easy to implement in 
our specific framework. The Maxwell stress density is
\begin{equation}
{\cal T}^{\rm H}_{\alpha\beta}({\bf r})=-{1\over 4\pi e^2}
[E_{\alpha}({\bf r})E_{\beta}({\bf r})
-{1\over 2}\delta_{\alpha\beta} E^2({\bf r})],
\label{hartree}
\end{equation}
where $E_{\alpha}$ is the electric field
\begin{equation}
E_{\alpha}({\bf r})=-\nabla_{\alpha}\int d{\bf r}'\>{\overline{\rho}({\bf r}')
\over |{\bf r}-{\bf r}'|}.
\label{campoel}
\end{equation}
To cure the coulombic divergences in
${\cal T}^{\rm H}_{\alpha\beta}({\bf r})$,
the  total charge density
$$\overline{\rho}  \equiv \rho^{\rm ion} + \rho$$ 
should be used in Eq. (\ref{campoel}),
rather than the electronic charge density $\rho$. The 
ionic charge is represented by a sum of ion-centered
Gaussians as
\begin{equation}
\rho^{\rm ion}({\bf r})=-\sum_j{Z_j \over \pi^{3/2}R_c^3}
e^{-|{\bf r}-{\bf R}_j|^2\over R_c^2},
\end{equation}
where ${\bf R}_j$ and $Z_j$ are ionic positions and charges, respectively,
and $R_c$ the gaussian radius. In this way ${\cal T}^{\rm H}_{\alpha\beta}$ 
contains the electron-electron interaction term,
\begin{equation}
-{1\over 4\pi e^2}[E^{\rm e-e}_{\alpha}({\bf r})E^{\rm e-e}_{\beta}({\bf r})
-{1\over 2}\delta_{\alpha\beta} {E^{\rm e-e}}^2({\bf r})],
\end{equation}
and the ion-ion interaction, 
\begin{equation}
-{1\over 4\pi e^2}[E^{\rm ion}_{\alpha}({\bf r})E^{\rm ion}_{\beta}({\bf r})
-{1\over 2}\delta_{\alpha\beta} {E^{\rm ion}}^2({\bf r})],
\end{equation}
plus an unphysical term due to the fictitious interaction between 
electronic and Gaussian-ion charge density, to be taken care of
below (see Eq. (\ref{locpp2})). This procedure ensures charge
neutrality within the volume bounded by bulk layers, besides that of
the  whole simulation cell. 
The crucial requirement that  the stress density integral
over a bulk layer inside, {\it e.g.}, a surface 
slab to be equal to the bulk stress,
is then satisfied. The electric field can be calculated in {\bf
G}-space and then carried to real space where ${\cal
T}^{\rm H}_{\alpha\beta}({\bf r})$ is easily   
evaluated.

A further contribution to the ion-ion interaction arises from the strain 
derivative of the ionic charge density. In real space, this Ewald-like
term reads
\begin{equation}
{\cal T}^{\rm Ewald}_{\alpha\beta}({\bf r})=
-{R_c^2\over 2}\nabla_{\alpha}\rho^{\rm ion}({\bf r})E^{\rm
ion}_{\beta}({\bf r}).
\label{ewald}
\end{equation}

For a purely coulombic system, the previous formulas  are all that
is needed to construct the stress density.
However, a practical implementation for the  plane-wave	%
 pseudopotential method calls for several extensions, due mostly to
the non-coulombic contributions of pseudopotential. 
The local pseudopotential stress contribution is
\begin{equation}
T^{\rm e-i}_{\alpha\beta}=\int d{\bf r}\> \rho({\bf r})
\sum_j {\partial \over \epsilon_{\alpha\beta}}V^{\rm loc}_j({\bf r}-{\bf R}_j),
\label{locpp}
\end{equation}
where $V^{\rm loc}_j$ is the local part of pseudopotential sited on atom $j$. 
From $\partial |{\bf r}|/\partial
\epsilon_{\alpha\beta}=r_{\alpha}r_{\beta}/|{\bf r}|$, it follows that
\begin{equation}
T^{\rm e-i}_{\alpha\beta}=\sum_j\int d{\bf r}\> \rho({\bf r})
{V'}^{\rm loc}_j({\bf r}-{\bf R}_j){(r-R_j)_{\alpha}(r-R_j)_{\beta}
\over |{\bf r}-{\bf R}_j|},
\label{locpp1}
\end{equation}
where ${V'}^{\rm loc}_j(x)=\partial V^{\rm loc}_j(x)/ \partial x$.
We define the corresponding stress density as
\begin{equation}
{\cal T}^{\rm e-i}_{\alpha\beta}({\bf r})=\rho({\bf r})\sum_j
{\partial (V^{\rm loc}_j({\bf r})
+ V^{\rm ion}_j({\bf r}))\over \partial \epsilon_{\alpha\beta}}
\label{locpp2}
\end{equation}
where the potential  $V^{\rm ion}_j$ generated by the
 the Gaussian ionic charges is added to compensate the same
contribution  present in Eq. (\ref{hartree}), and to cancel the
coulombic  
divergences in ${\cal T}^{\rm e-i}$.
The strain derivative of local pseudopotential in {\bf G}-space is
(for brevity we use the same symbol for a quantity in {\bf G}- and
{\bf r}-space)
\startlongequation
\begin{eqnarray}
{\partial V^{\rm loc}_j\over \partial \epsilon_{\alpha\beta}}({\bf G})=
\sum_{{\bf G}'\neq 0} S_j({\bf G}')\left [\left ({\partial V^{\rm loc}_j(G')
\over \partial {G'}^2}-{4\pi e^2 \over {G'}^4}\rho^{\rm ion}_j(G')\right)
\>2\>{G'}_{\alpha}{G'}_{\beta}+\left(V^{\rm loc}_j(G')+
{4\pi e^2 \over {G'}^2}\rho^{\rm ion}_j(G)\right)\delta_{\alpha\beta}\right]
\label{pippo6}
\end{eqnarray}
\stoplongequation
for all ${\bf G}\neq 0$, whereas the ${\bf G}=0$ contribution is
\begin{equation}
{\partial V^{\rm loc}_j\over \partial \epsilon_{\alpha\beta}}(0)=
\left[ \alpha_j-{\pi e^2 Z_j R^2_c \over \Omega}\right].
\delta_{\alpha\beta}
\end{equation}

 Eq. (\ref{pippo6}) is derived using the fact that a positive
strain ${\hat{\epsilon}}$ in {\bf r}-space corresponds to linear order
to ${\bf G}'=(1-\hat{\epsilon})\,{\bf G}$ in {\bf G}-space, thus 
\begin{eqnarray}
{\partial V^{\rm loc}_j ({\bf G})\over \partial \epsilon_{\alpha\beta}}=
-{\partial V^{\rm loc}_j({\bf G})\over G^2} 2 \,G_{\alpha} G_{\beta}.
\end{eqnarray}
The {\rm G}-space derivative of $V^{\rm loc}_j$ is evaluated
numerically.

The non local pseudopotential term only acts in the core regions,
where spatial resolution  is largely arbitrary. Therefore,
we choose it to be a superposition of ionic-site centered contributions,
\begin{equation}
{\cal T}^{\rm nl}_{\alpha\beta}({\bf r})=\sum_j\>P^j_{\alpha\beta}
\delta({\bf r}-{\bf {\tau}}_j),
\label{nlstress}
\end{equation}
where 
stress
\begin{equation}
P^j_{\alpha\beta}=\sum_{\nu{\bf k}}f_{\nu{\bf k}}\omega_{\bf k}
\langle\psi_{\nu{\bf k}}|{\partial V^{NL}_j\over \partial
\epsilon_{\alpha\beta}} 
|\psi_{\nu{\bf k}}\rangle
\end{equation}
is the $j^{th}$-ionic site contribution.
The most general form of a fully non local pseudopotential can be 
written as a non-diagonal projector
\begin{equation}
V^{NL}_j=\sum_{nm}D_{nm}^j|\beta_n^j\rangle\langle\beta_m^j|,
\end{equation}
where $n$, $m$ are sets of atomic quantum numbers. In the case of ordinary 
Kleinman-Bylander pseudopotentials, the matrix $D_{nm}^j$ is diagonal and 
its elements are constant, thus the corresponding stress is given by the 
strain derivatives of the projector functions $\beta_n^j$. 
In plane waves, we have in general
\startlongequation
\begin{eqnarray}
P^j_{\alpha\beta}=
{P^j_{\alpha\beta}}^{us} + 
\sum_{{\bf k},{\bf G},{\bf G}',\atop {\nu},n,m}f_{\nu {\bf k}}
\>\omega_{\bf k}D_{nm}^j e^{-i({\bf G}-{\bf G}')\cdot{\bf R}_j}
\> \psi_{\nu{\bf k}}^*({\bf G})
{\partial [\beta_n^j({\bf k}+{\bf G})\> {\beta_m^j}^*({\bf k}+{\bf G}')]
\over \partial \epsilon_{\alpha\beta}}
\psi_{\nu{\bf k}}({\bf G}'),
\end{eqnarray}
\stoplongequation
where
\begin{eqnarray} 
{\partial \beta_n^j({\bf k}+{\bf G})\over\partial \epsilon_{\alpha\beta}}=
{\partial \beta_n^j({\bf k}+{\bf G})\over\partial (k+G)_{\alpha}}\,
 (k+G)_{\beta}.
\end{eqnarray}
The term 
${P^j_{\alpha\beta}}^{us}$ is only  non-zero if
  ultrasoft pseudopotentials\cite{van} are used.
In that case, $D^j_{nm}$ is non diagonal and contains a
charge-dependent term,
\begin{equation}
D^j_{nm}=D^0_{nm}+\int d{\bf r}\>Q_{nm}^j({\bf r}-{\bf R}_j)
[V^{\rm loc}({\bf r})+V^{\rm Hxc}({\bf r})],
\end{equation}
where $V^{\rm Hxc}$ is the screening (Hartree plus exchange-correlation) 
potential. The strain derivative of $[V^{\rm loc}({\bf r})+V^{\rm Hxc}({\bf r})]$
was already taken into account in Eq. (\ref{locpp2}), so 
the strain derivative of $D^j_{nm}$ only contributes the  term
\begin{eqnarray}
{P^j_{\alpha\beta}}^{us}\!=
\sum_{n,m,{\bf G}}B^j_{nm}e^{-i{\bf G}\cdot{\bf R}_j}
{(V^{\rm loc}}+V^{\rm Hxc})^*({\bf G}){\partial Q_{nm}({\bf G})\over 
\partial \epsilon_{\alpha\beta}},\nonumber
\end{eqnarray}
with
\begin{equation}
B^j_{nm}=\sum_{\nu{\bf k}}f_{\nu{\bf k}}\>\omega_{\bf k}
\langle\psi_{\nu{\bf k}}|\beta_n^j\rangle 
\langle {\beta_m^j}|\psi_{\nu{\bf k}}\rangle.
\end{equation}

The formulation of the stress density is now completed. The
 final expression reads
\begin{eqnarray}
{\cal T}_{\alpha\beta}({\bf r}) & =& 
{\cal T}^{\rm kin}_{\alpha\beta}({\bf r}) +
{\cal T}^{\rm H}_{\alpha\beta}({\bf r}) + 
{\cal T}^{\rm xc}_{\alpha\beta}({\bf r}) + 
\label{finale} \\&&
{\cal T}^{\rm Ewald}_{\alpha\beta}({\bf r}) +
{\cal T}^{\rm e-i}_{\alpha\beta}({\bf r}) +
{\cal T}^{\rm nl}_{\alpha\beta}({\bf r}), \nonumber
\end{eqnarray}
the involved terms being given respectively by
Eqs. (\ref{kin1}),
(\ref{xc}),
(\ref{hartree}),
(\ref{ewald}),
(\ref{locpp2}), and (\ref{nlstress}).
Operatively, all  terms (except exchange-correlation)  are 
first calculated in {\bf G}-space and then Fourier-transformed
 to {\bf r}-space, with a 
computational load comparable to that needed to calculate the 
macroscopic stress tensor. 
In order to accurately evaluate the 
stress density and the stress itself,
 a well-converged selfconsistent charge 
density is generally needed, due to the  non-variational character of 
the stress,  and its sensitivity to the details of the charge density. 

\subsection{Averages and stress density handling}
A practical point concerns how to handle and  visualize
the stress density (more discussion and examples
are given in Sec. \ref{applica}). When dealing with surfaces and
interfaces,  appropriate averages of the stress density  can be taken,
that  contain all the relevant information. If $\hat{z}$  is the
surface normal,  the {\it planar average} is defined as
\begin{eqnarray}
\overline{\cal T}(z) &= &{1\over A}\int dx\, dy\>{\cal T}({\bf r})\\ &=&
\delta_{G_x,0}\delta_{G_y,0}\sum_G\>{\cal T}({\bf G})e^{-iG_z\cdot z}.
\nonumber
\end{eqnarray}
\noindent
A typical planar average (see Fig. \ref{allumi}) is strongly
oscillating, with negative peaks at the ionic positions (corresponding
mostly to attractive ion-electron contributions) and positive peaks in the
interstitial region (corresponding mostly to kinetic compressive
stress). Deviations from bulk-like behavior, signaling surface-induced
stress changes, can be directly inspected. 

A further convenient
way of analyzing $\cal{T} ({\bf r})$ is its {\it macrosopic
 average} \cite{bbr}  over a period of length $d$, and
defined as
\begin{equation}
\overline{\overline {\cal T}}(z)=
{1\over d}\int_{z-d/2}^{z+d/2} \!dz'\,\, \overline{\cal T}(z').
\end{equation}
This filtering operation eliminates  oscillations of period $d$, and 
can be generalized\cite{bbr2} to account for several superimposed
oscillating components. 
If $d$ is, {\it e.g.}, the interlayer distance in a
metal slab,  then $\overline{\overline {\cal T}}(z)$ is constant in
the bulk region of the slab, and its value equals the 
bulk stress; in the vacuum region $\overline{\overline{\cal{ T}}}$
vanishes by construction. Any deviation from these constant values
in the simulation cell quantify surface-related
stress changes. In {\bf G}-space the macroscopic average reads
\begin{equation}
\overline{\overline {\cal T}}(G_z)=
{2\over d}\>{\overline{\cal T}(G_z)\over G_z}
\sin\left({d\>G_z \over 2}\right),
\end{equation}
where $\overline{\cal T}(G_z)={\cal T}(0,0,G_z)$.
As discussed in detail in Ref. \onlinecite{cm},
the macroscopic average is gauge-independent by construction.

\subsection{Gauge dependence}
\label{gauge}

A major concern  about the applicability of the
stress density (as in the energy density case)
is its formal gauge dependence, showing up e.g.
in the  two distinct forms Eq. (\ref{kin}) and
(\ref{kin0}) of the kinetic stress contribution.
Gauge dependence, however,
does not affect  the three classes of
applications we are interested in here, which involve
 integrated stresses, macroscopic averages of the stress density,
and changes in the planar averages thereof.
\begin{enumerate}
\item By definition, the gauge indeterminacy does not
 affect observables such as the {\it  integrated} stress, 
hence it does not prevent  the usage of the stress density to extract, say, 
integrated surface stresses  in an
improved or technically safer fashion (Sec. \ref{applica}).
\item As discussed in detail in Ref.\onlinecite{cm}, the
 {\it macroscopic} average of the  stress density is
gauge-independent by construction, since   gauge functions
are lattice-periodic, and hence are filtered out by
 macroscopic averaging.
\item Changes in stress density in response to 
perturbation are gauge-independent.
\end{enumerate}
The proof of the last statement is as follows.
Let     ${\cal T}[\rho({\bf r})]$
and ${\cal T}'[\rho({\bf r})]$ be two functionals of the density
whose integral is the same:
\begin{equation}
\int d {\bf r}\>{\cal T}[\rho({\bf r})]=
\int d {\bf r}\>{\cal T}'[\rho({\bf r})].
\label{eq1}
\end{equation}
[An interesting choice is the two functionals  being
 the stress densities  with  the two kinetic
gauges Eqs. (\ref{kin}) and (\ref{kin0}).] 
Under  a generic perturbation on the Hamiltonian,
\begin{equation}
H_{\lambda}=H+\lambda V,
\end{equation}
  ${\cal T}$ becomes to linear order
\begin{equation}
 {\cal T}={\cal T}[\rho_0]+
\left({\partial {\cal T}\over\partial \rho}\right)_{\!\!\!\rho_0} 
\>\delta\rho_{\lambda}+...
\label{ark}
\end{equation}
As the unperturbed
integrated values are equal,
\begin{equation}
\int d {\bf r}\>{\cal T}[\rho_0]=
\int d {\bf r}\>{\cal T}'[\rho_0],
\end{equation}
substituting  Eq. (\ref{ark}) into Eq.(\ref{eq1})  we
get
\begin{eqnarray}
0 &=&  \int d {\bf r}\> \left({\cal T}[\rho_0] - {\cal T}'[\rho_0]\right) = 
\nonumber\\
&=&\int d {\bf r}\>
{\delta \rho}_{\lambda}
 \left[\left({\partial {\cal T}'\over\partial \rho}\right)_{\!\!\!\rho_0}
- \left({\partial {\cal T}\over\partial \rho}\right)_{\!\!\!\rho_0}\right]
\end{eqnarray}
The density variations are arbitrary, so we conclude that
\begin{equation}
\left( {\partial {\cal T}\over\partial \rho}\right)_{\!\!\!\rho_0}=
\left( {\partial {\cal T}'\over\partial \rho}\right)_{\!\!\!\rho_0}.
\end{equation}
Therefore, two density-functionals obeying Eq. \ref{eq1}
have identical variations under
 external perturbation, to linear order with respect
to the density. We then
rest assured that stress density  changes
are gauge-independent to linear order.
This puts on firm ground the  kind of application
exemplified  in Section \ref{micro}, which  rely on the 
analysis of changes in stress density planar averages;
typically, one will compare e.g.
 planar averages in different systems and instances  instead 
of analyzing differences (which is
justified, as just proven): the reason for this is simply the
better  pictorial rendering of the former approach.

As to stress densities in themselves (more precisely any 
non-macroscopically  averaged ones), 
they are unavoidably non-unique.  
In practice, however, we found
 numerically very similar results using
 both the possible choices for the    kinetic stress density;
in addition, the spatial behavior of e.g. planar averages
 agrees closely with  physical intuition (as discussed in Sec. 
\ref{applica}).

\section{Applications of the stress density}
\label{applica}
\subsection{Non-polar surfaces: tests on Al (111)}
\label{nps}

To illustrate the basic characteristics of the stress density 
we first discuss
 the (111)--$1\times 1$ surface of Al in the ideal (i.e. unrelaxed)  
structure. Calculations employed a 9-layers slab,
norm-conserving pseudopotentials, and  well converged
plane-wave and k-point sets.  In our formalism a negative stress is
tensile, {\it i.e.} it indicates that reducing the interatomic distance
is energetically favorable; a positive stress is compressive --
that is, the system prefers to be stretched. 
The stress densities considered in the following are planar  
or macroscopic averages of the in-plane components 
${\cal T}_{xx}({\bf r})={\cal T}_{yy}({\bf r})$ (almost all the systems 
considered hereafter are isotropic in the surface plane):
indeed, only the in-plane stress matters, 
since the orthogonal component is eliminated by relaxations.

\begin{figure}
\narrowtext
\epsfxsize=8.5cm
\centerline{\epsffile{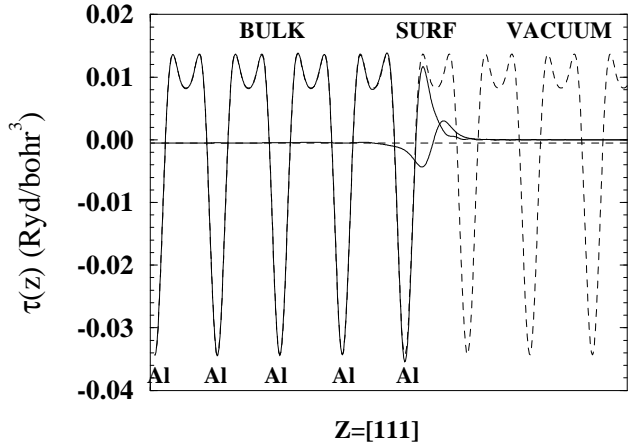}}
\caption{Stress density of the unrelaxed Al (111) surface.
Solid lines are planar and macroscopic averages for
the surface slab, dashed lines same quantities for the bulk 
in the (111) slab symmetry (see discussion in text).
\label{allumi}}
\end{figure}

In Fig. \ref{allumi} four curves are presented: two of them, the 
solid lines, are the planar and macroscopic averages of 
the stress density for the surface; the other two, the 
dashed lines, are the same quantities for the  Al bulk in 
(111) symmetry (obtained  filling up  with atoms
the vacuum region between repeated slabs). 
The comparison serves to  better visualize changes due to 
the surface formation. Thanks  to  symmetry  along [111], 
only half of the simulation cell is shown in Fig. \ref{allumi}. 

As mentioned earlier, the planar averages are characterized
 by oscillations having the period of the interlayer distance
in the bulk region. Negative peaks correspond to the layer positions by
construction, and mostly store the tensile contribution  
due to the electron-ion interaction, whereas the positive peaks 
located between layers are dominated by compressive kinetic and 
coulombic terms. The surface planar average drops to
zero on entering the vacuum region, and it is informative to compare
it with the bulk planar average near the surface. 

The macroscopic average of the surface is constant  in the bulk
region and zero in  vacuum,{\it  i.e.} it is not constant only
in the region where surface effects arise. Thus, it is enables one
to  identify the  boundaries of the surface region. In the present
case, for instance,  the perturbation produced by surface formation
extends roughly over one interlayer  distance at  the solid-vacuum
interface.  The macroscopic average for the bulk is everywhere
constant, and it equals  the surface macroscopic average in the bulk
region. This common value of the two macroscopic averages in the bulk
region, equaling  by construction the bulk stress, is close but not
equal to zero. 

This is a first important capability  of the stress density.
It is a common feature of  stress calculations that the
stress be non-zero 
even though the lattice constant corresponds to  theoretical equilibrium
(hence in principle to zero stress). Indeed, energy minimization is
usually performed at fixed plane-wave cutoff energy ({\it i.e.}
 variable number  of plane-waves), whereas the stress is implicitly
calculated at fixed  number of plane waves.\cite{feib,van2}
This bias, specific of the plane-wave methodology, can be eliminated
using an absolutely  converged basis set, but that is  highly
impractical especially when dealing with large systems. Thus, 
 the  integrated  surface stress should be obtained
substracting $n$-times the bulk contribution 
(if $n$ is the number of  atoms, or more generally of
formula units, in the slab) from the total stress 
as
\begin{equation}
T^{\rm surf}_{\alpha\beta}={1\over 2}\left[T^{\rm slab}_{\alpha\beta}-
n\>T^{\rm bulk}_{\alpha\beta}\right]
\label{surfst}
\end{equation}
where factor $1/2$ accounts for the two slab  surfaces.
 As for the surface energies,\cite{fm} this procedure is justified in general
to  the extent that the finite-size basis error is proportional to the
number of atoms, with the additional proviso that
 the error per atom does not change significantly
for small changes in volume per atom.\cite{van}

The stress density provides a
straightforward evaluation of the spurious bulk stress,
 requiring no separate calculations, because 
 the spurious bulk stress per atom is just the value of the
macroscopic stress density average in the bulk region.
Calculating the spurious stress in this fashion not only 
saves computing time, but also gives more accurate results,
since use of the bulk stress calculated in the bulk unit cell 
in Eq. (\ref{surfst})  introduces  numerical errors beyond the
basis-set error, due {\it e.g.} to inequivalent k-points.  Another
source of error is  surface stress anisotropy (such as it occurs in
most reconstructions): the spurious bulk stress  is also expected to
be anisotropic, a feature that cannot be but missed in the bulk unit
cell.

Coming now to the shape of the macroscopic stress density average,
 we see that the surface formation causes an excess of tensile stress
just below the surface-vacuum interface, partially compensated
by an overshot of  compressive stress outside the interface. The
resulting total surface  stress is slightly tensile.  
The presence of tensile stress is a common feature of most metal 
surfaces,\cite{fms,feib1,needs}
 and its origin is easily understandable
for $sp$ metals\cite{needs}  in terms of
electron spill-out into vacuum, resulting in a reduction of
 the kinetic  compressive stress in the 
near-surface region (the interpretation is slightly more
involved\cite{fms} for transition metals). This tensile stress remains built-in
at the surface,  since surface atoms are constrained to their
positions by  the substrate potential, and in-plane contraction would
require the formation of (typically costly) defects to be
topologically possible. 

We close this Section with an example of layer-by-layer resolution
(LBL) of the stress.  In Table  
\ref{fit_tab} we list the LBL of the total stress for Al (111),
i.e. the integrals of the planar stress density average over one
interplanar distance. The 15 ``geometrical'' slab layers
(9 are occupied by atoms) are labeled --7 to 7, layer 0 being
 the bulk-like slab center, and layers +n and --n being
  equivalent by inversion symmetry along
$\hat{z}$. Layers --3 to +3 are bulk-like, having nearly
equal LBL value. Their average stress  (--0.743 eV) equals
to within 1 meV the bulk stress calculated in the bulk fcc. Inserting
this value into Eq. (\ref{surfst})  we obtain a value for the surface
stress (--0.58 eV/atom) unaffected  by numerical errors.   
The resulting surface stress is in good agreement with the results of
a  similar procedure reported by Feibelman.\cite{feib1} 

\subsection{Polar surfaces: GaAs (100) and (111)}
\label{ps}

As briefly mentioned earlier,  orientations for which cleavage
produces  inequivalent surfaces are a natural playground for the
stress (and energy) density.  We consider here two prototypical
cases, namely the (100) and  (111) surfaces of (zincblende) GaAs, 
which have been and still are the focus of intense
experimental
\cite{mess,weiss,biegel,notz} and 
theoretical
\cite{cm1,rcyb,mkps,kax,qmc,ohno,nofr}
 work.  These surfaces are  polar, 
and present  several (otherwise unpleasant)
 features making them suitable for a test of the
stress density. 

 For the (100) orientation,  each surface (Ga or As terminated) can
indeed be individually simulated in a symmetric slab,  so that the
energy of each surface can be calculated via total energy differences;
for testing purposes, however, we will also consider asymmetric slabs
where inequivalent surfaces are coupled.
For the (111) orientation, the surfaces are intrinsically different.
Either are they geometrically identical but terminated by a
different atom, or  terminated with the same atomic
species but with different geometries: the  (say) Ga-terminated 
surfaces of a (111)-oriented slab are necessarily one (111) and one
($\overline{1}\overline{1}\overline{1}$), and no symmetrization is
practicable in this case.  Using the energy density is then
indispensable \cite{cm1,rcyb,mkps} for a direct
calculation of the absolute surface energy of each surface,
 whereas only averages, or  differences to some reference system are 
accessible   by plain total energy calculations.   The procedure
 is not  straightforward, since it  requires counting the atoms  that
belong to a given surface; this can be  troublesome for GaAs
(111), and arguments based on cell symmetry adaptation\cite{cm1} or
space partitioning\cite{rcyb} have to be  employed to  complement the
use of the energy density.  Nevertheless, the energy density is more
computationally convenient to study this surface than the total energy
is. 

Analogous considerations hold for the stress density, with the significant
difference that there is no need to count surface atoms to evaluate
the surface stress, and one need only making sure that  the bulk
structure is at theoretical equilibrium and the cutoff energy is large
enough to render negligible the finite-size basis error.

There are other difficulties  related to the presence of unequivalent
surfaces. The two workfunctions are not lined up, and fictitious
electric fields  are produced in the vacuum region.
 Furthermore, as-cleaved GaAs (100) and (111) are 
metallic, because  partially filled dangling-bond
surface states are present on Ga-  as well as As-terminated
surfaces. As a consequence, a  sizable charge transfer between the
surface bands  occurs across the slab;  this artificial charge 
flow can be avoided  resorting to passivation with, typically,
fractionally charged H atoms.   In reality these surfaces
reconstruct into complex atomic arrangements to restore neutrality
and remove metallicity;
however, we decided to consider  here only the unreconstructed
and  unpassivated surfaces, since they represent a  technically
intricate and critical case of supercell surface calculation
 if there ever was one, and are ideally suited as a testing
ground  for the stress density.   We left the surfaces unrelaxed at
the theoretical lattice constant. 

 Ga-terminated (Ga-t)  and As-terminated
(As-t) GaAs (100) can be represented individually within symmetric
slabs. In this way, essentially no charge transfer across the slab
occurs, barring the small residual interaction between surfaces
which splits  the  (in principle degenerate) surface bands. 
On the other hand, the Ga-t and As-t surfaces
 can also be made to appear simultaneously  in the same cell.
This configuration is undesirable in  general as it leads to sizable
charge transfer, but it is useful as a test case; surface stresses
 obtained in the two symmetric (charge transfer unaffected) slabs can be
directly compared with those of the  coupled-surfaces asymmetric slab.
For these $1\times1$ surfaces we 
use 12 (plus 6 vacuum) layers for the asymmetric slab, 11 (plus 7) layers 
for the symmetric one. Ultrasoft pseudopotentials \cite{van} are
employed. 
The macroscopic averages are obtained by averaging over a distance of two 
interlayer spacings.

\begin{figure}
\narrowtext
\epsfxsize=8cm
\centerline{\epsffile{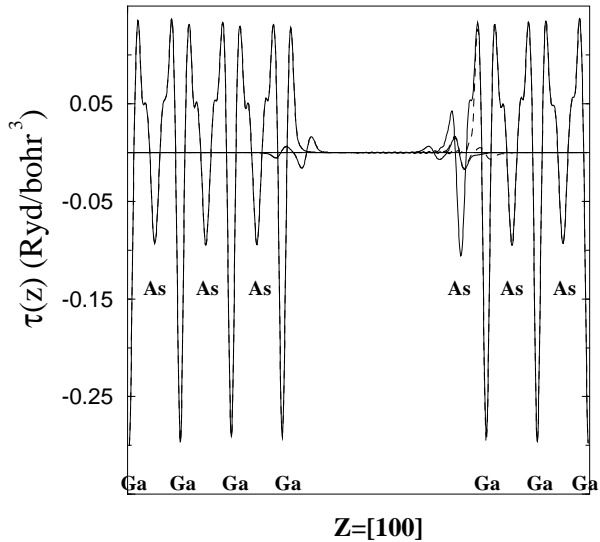}}
\caption{Planar and macroscopic averages of the stress density for
GaAs (100). Solid lines refer to the (asymmetric) slab containing 
both Ga-t and As-t surfaces, dashed lines to the symmetric slab for the 
Ga-t surface. Macroscopic averages vanish in the bulk and in the vacuum,
and their integrals over the surface regions give the surface stresses.
Solid and dashed lines are  distinguishable only in the region of the 
As-t surface.
\label{pippo1}}
\end{figure}

In Fig. \ref{pippo1} we show planar and macroscopic averages of the stress 
density in the asymmetric slab (solid lines), and the same quantities in
the symmetric Ga-t slab (dashed lines).  The macroscopic average
inside the slab show  that the bulk stress is zero to numerical
accuracy,  so that the surface stress is given directly
by  stress density integrals over one half of the
slab. [Because  the surfaces are non-stochiometric, Eq. (\ref{surfst})
cannot be used to eliminate the spurious bulk stress; we therefore had
to use  a sufficiently converged (60 Ryd cutoff)  plane-wave basis so
as to achieve a negligible residual bulk stress.]  In the bulk region
the averages for the symmetric and asymmetric slabs are
indistinguishable, indicating that in both structure the inner slab
region is insensitive (stress-wise)  to the surface. At the  Ga-t
surface the  various averages match almost perfectly, indicating that 
 the slabs are large enough to entirely decouple (stress-wise)
the  surfaces (mind that in one case the other surface is identical,
in the other it is As-t). As expected, the calculated charge transfer
is much larger for the asymmetric  (0.02 electrons per cell, or 5
mC/m$^2$)  than for the symmetric slab (0.5 mC/m$^2$). The surface
stress per surface cell  obtained integrating the stress density  upon
the half slab containing the Ga-t surface is --1.114 eV/atom per surface
cell for 
the asymmetric slab, and --1.148 eV/atom for  the symmetric one. Thus the
error due to charge transfer is about 3\%, 
and  well below  0.1 eV per surface
cell, which is similar to that of   half-slab integrals
of the energy density in Ref.\onlinecite{cm}, where an  
analogous test was performed.
Notice that for these unrelaxed structures the surface effects are
very much localized on the surface atoms, and the bulk features are
substantially recovered just one layer below the surface. Also, not
surprisingly, the surface stress is tensile.

\begin{figure}
\narrowtext
\epsfxsize=8cm
\centerline{\epsffile{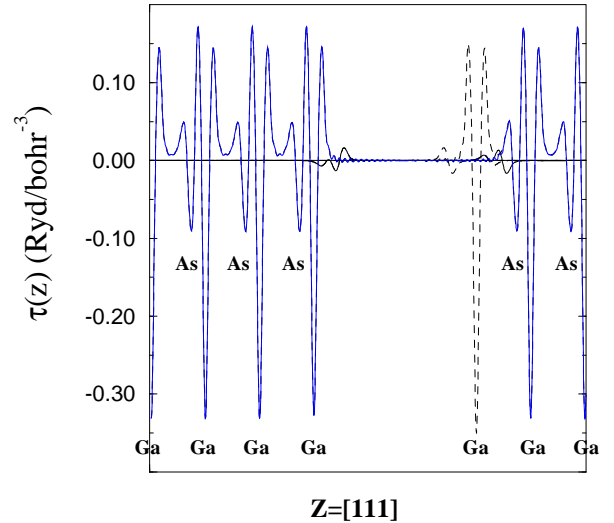}}
\caption{Planar and macroscopic averages of stress density for 
GaAs (111). Solid lines refer to the slab with both Ga-t and As-t
surfaces, dashed lines to that with (111) and 
($\overline{1}\overline{1}\overline{1}$) Ga-t surfaces. 
\label{pippo2}}
\end{figure}

As already mentioned, for GaAs (111) there is no way to 
set up a symmetric slab containing  only one kind of surface,
since (111) and ($\overline{1}$$\overline{1}$$\overline{1}$) surfaces 
are intrinsically different, and not related by any symmetry operation.
We then we consider a comparison between the structure with both Ga-t
and As-t (111) surfaces, and that one with Ga-t (111) and
($\overline{111}$) surfaces  (Fig. \ref{pippo2}). The solid lines
represent planar and macroscopic  averages of the slab with chemically
different surfaces, the dashed ones  are the analogous quantities for
the Ga-t surface slab. Both the structures are now affected by an
equal spurious charge transfer $\sim$ 0.04 electrons per cell;
this is not unexpected since  electron counting assigns 
in both cases 
the same occupation numbers to partially-filled surface bonds, that is, 
3/4 on Ga-t (111) and 1/4 on As-t (111) and Ga-t ($\overline{111}$).
The comparison of the half-slab stresses containing the same Ga-t surface,
{\it i.e.} the Ga-t surface stress, show an almost perfect agreement: we
obtain --0.515 eV for the same-ion surface slab, and --0.508 eV for
the different-ion surface slab. This is due both to the large number of 
bulk layers ensuring that there is no interaction between the surfaces 
in the slab, and to the (partially fortuitous)  equality of charge
transfer in the two slabs. 

\subsection{Ge/GaAs (100) and (110) interfaces}
\label{inter}

In this Section we present calculations on  Ge/GaAs interfaces.
For interfaces, the computing effort saved by using 
space-resolved quantities can be even higher than for the
surfaces. First,  the considerations of Sec. \ref{ps} for polar
surfaces apply in full also to inequivalent interfaces.  Second,
  the simulation cell sizes are generally larger than for surfaces,
hence the ability of avoiding separate calculations is very much
welcome. Third,  the relevant quantities entangled in the integrated
stress (or energy)  are at least three, namely the interface excess
stress and the bulk stress of the two bulk phases,  so that more
separate bulk calculations  are needed within the usual  approach
than in the surface case.  

\begin{figure}
\narrowtext
\epsfxsize=8cm
\centerline{\epsffile{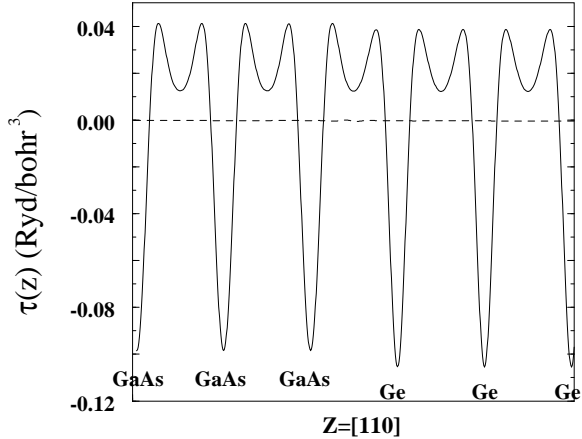}}
\caption{Planar average (solid line) and macroscopic average (dashed line) 
of the stress density for the (110) Ge/GaAs interface. The presence of the
interface is barely visible in the macroscopic average, since the
interface is non polar and only very slightly mismatched. Only 
half slab is shown. 
\label{interf1}}
\end{figure}

As an example of stress density application we first present the case of
the non-polar  abrupt  $1\times1$  (110) Ge/GaAs interface. 
In Fig. \ref{interf1}, planar and macroscopic  averages of the
stress density are shown for a (GaAs)$_5$ (Ge)$_5$ slab. 
The in-plane lattice constant is fixed at the theoretical  value for
bulk GaAs; since	the equilibrium lattice constant of  bulk Ge 
is slighty larger,  the Ge film is under tensile stress.
The structure is unrelaxed, and the interlayer distances are those of
bulk GaAs. 
[Note that
 these interfaces are stoichiometric, hence Eq. (\ref{surfst})
can be used to get rid of the spurious bulk stress.
The  highly  converged plane-wave basis used for the
polar surfaces is not mandatory any more, and 
the cutoff was lowered to 25 Ryd. This produces an appreciable
residual stress in GaAs.]
The stress induced by the interface is barely
visible in Fig. \ref{interf1}; only a small dip in the
macroscopic average can be distinguished between  the interface
 GaAs and Ge layers.   For a non polar junction this is quite reasonable,
 since the charge transfer across the interface is too small and
interface-localized to produce a significant excess  stress.
Integrating the planar (or macroscopic) average over the bulk regions
we obtain $T_{\rm bulk}^{\rm GaAs}$=--0.94 eV/atom and $T_{\rm
bulk}^{Ge}$=--1.52 eV/atom. 
The resulting interface stress of  $T^{int}$= 0.02 eV/cell $\sim$ 1 meV eV/\AA\,
 is practically vanishing.
Clearly this nearly-matched interface is a test case where the stress due 
to the interface formation does not play  any major
role. Nevertheless,  even in this case the power and semplicity
of the stress density  approach is evident.

\begin{figure}
\narrowtext
\epsfxsize=8cm
\centerline{\epsffile{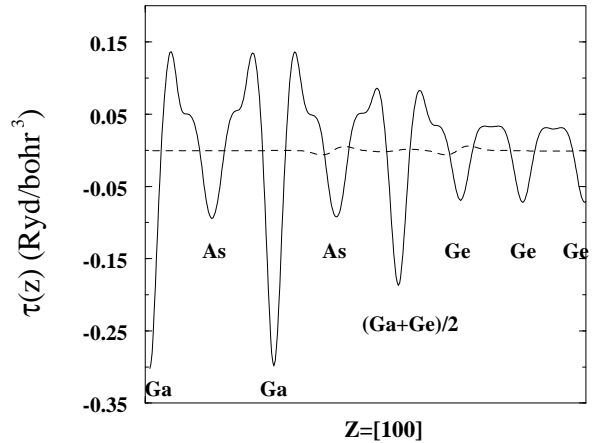}}
\caption{Planar (solid line) and macroscopic (dashed line) average of the 
stress density for the mixed $2\times 1$ (100) Ge/GaAs interface. 
GaAs and Ge are interfaced by a mixed  Ge/Ga layer.
By symmetry, only one interface is shown.
\label{interf2}}
\end{figure}

A more interesting case of interface-induced stress is illustrated
in   Fig. \ref{interf2}, where we show planar and macroscopic
averages of the stress  
density for the mixed $2\times 1$  (100)  Ge/GaAs interface.  This 
 polar interface is simulated with two bulk regions comprising 5 Ge
and  5 GaAs layers, matched  by  a mixed Ga-Ge layer
 (12 layers and 24 atoms in total). The slab contains an equal  number
of Ga and As atoms, so that  the bulk GaAs contribution can be
unambiguously  subtracted out. It is also easily  verified
by the electron-counting rule\cite{abh}  that no metallic  states are 
present at  the interface, which is in fact a reasonable candidate as
the  actual interface structure.

The theoretical bulk GaAs  in-plane lattice constant is imposed to the
structure; the value of the macroscopic average in the bulk GaAs
region represents just the  finite-size basis error, whereas the
bulk Ge region  is also subject to  strain. The bulk stresses
are $T_{\rm bulk}^{\rm GaAs}$=--0.95 eV/atom,  and $T_{\rm
bulk}^{\rm Ge}$=--1.61 eV (the small differences from the values obtained
 in the (110) Ge/GaAs interface is due to k-point sampling). 
The interface region extends  (stress-wise) 
over  three layers,  the  mixed interface layer and the two adjacent
 to that. Outside this restricted region, the  macroscopic average quickly
recovers the equilibrium bulk value.  The interface stress is 0.21
eV/atom, i.e. a condition of small  compressive stress
induced by the Ge epilayer.

\subsection{Microscopic analysis based on the stress density}
\label{micro}

So far we have been giving examples of applications of the stress 
density to the calculation of macroscopic stress  in cases where direct
calculations would be  troublesome. Actually, however,  the stress
density appears to be most useful in analyzing
 processes   driven or simply influenced by the
stress, as a kind of ``stress microscope''. 
Reconstruction, adsorption, and growth are some of the
 phenomena in which stress can play a relevant role,
  possibly be a driving force, or  simply function as
 a useful indicator of the ongoing processes.\cite{ibach,fms} 
The integrated stress, though, is sometimes  of limited use,
and occasionally misleading. While it does tell us whether the
application of strain will be  
energetically favorable, it easily misses the microscopic 
details, and cannot answer several legitimate questions such as: How
far does the perturbation propagate below a surface or across an interfaces?  
How is the stress redistributed\cite{me1} into the perturbed region?
What interplay of tensile and compressive  contributions 
 \cite{me2} produces the resulting integrated stress ? 
Sometimes answering these questions is fundamental to attain a
 consistent qualitative view of a specific process, and indeed
the stress density has been usefully applied in this fashion to the study
of surface reconstructions.\cite{me1,me2} Here, in particular, we
discuss briefly the hex reconstruction  of Ir  (100),
an apt example of what we mean by a microscopic 
analysis performed using stress density. 

In this reconstruction, a  quasi-hexagonal buckled layer
(whereby 6 atoms take place into a 5$\times$1 surface area)
is formed on top of  the square (100) substrate. The increase 
in surface atomic density is expected to have a close relation to
 the surface stress, which  is tensile and very large
(--1.86 eV/1$\times$1 area) on the unreconstructed surface.
The easy guess (the ``stress hypothesis'')  would be that the
densifying reconstruction is driven 
by the  energy gained by partially or completely relieving the
stress of the pristine surface. This hypothesis was indeed 
 made plausible by comparing the stress-driven energy gain
with  model estimates  of the substrate-surface rebonding cost, based
{\it e.g.} on bond cutting arguments, or models of the defects needed
to create the   reconstruction. \cite{fms,sao} 

\begin{figure}
\narrowtext
\epsfxsize=7cm
\centerline{\epsffile{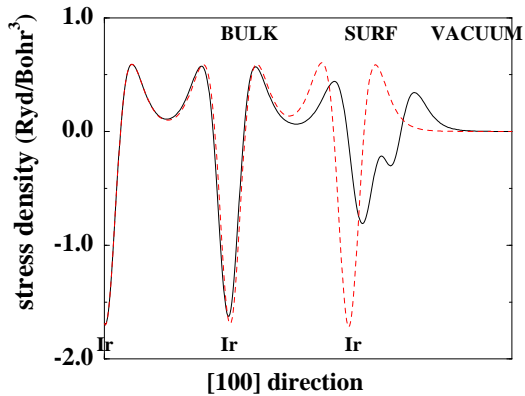}}
\caption{Planar average of the stress density for 
 unreconstructed (dashed line) and  5$\times1$-reconstructed 
(solid line) Ir (100).
The integrated stress is negative (i.e. tensile) for both the surfaces,
and larger in magnitude for the reconstructed one.
\label{5x1}}
\end{figure}

The validity of previous models
\cite{fms,needs,sao} does not depend on the value of the
stress after the reconstruction;  it is reasonable to assume, though,
 that the stress should decrease upon reconstruction,  fittingly
to its role of driving force. 
However, we found \cite{me2,noi} 
 that the surface tensile stress actually {\it increases} in
magnitude by about 20\%  
 or 0.3 eV/1$\times$1 area (on average on the
$\hat{x}$ and $\hat{y}$ 
stress  components of the anisotropic reconstructed surface)
on Ir (001). Similar
results are suggested by experiments \cite{ibach2} for Au (001).
These  new evidences on the integrated stress have increased the
confusion, leading some to conclude\cite{ibach2}
that the ``stress hypothesis'' is invalid, {\it i.e.} that 
stress is not the driving force to reconstruction,
and enhancing the mismatch between data and physical intuition.
It turns out that the application of stress density\cite{me2}
provides a natural explanation of the reconstruction mechanism,
reconciling all previous theoretical and experimental observations.

In Fig. \ref{5x1} the  stress density
planar average  is shown for
the (fully relaxed) reconstructed and unreconstructed (100)
surfaces. Upon reconstruction 
 the  outermost negative and two outermost positive peaks,
storing tensile and compressive contribution, respectively, are
strongly  reduced. The  surface layer relaxes outwards  by a 
huge  20\% 
 of the ideal interplanar distance, which is also reflected in the
outward shift of the tensile peak. The smearing of the latter
is a token of the large surface buckling. 
From these findings, the following picture 
readily emerges. The decreased interatomic
distance in the surface layer correlates with the severe
  reduction of the tensile stress 
within that layer (suppression of tensile peak). 
At the same time, the surface layer needs to relax
 outward to reduce its interaction with the substrate potential;
this displacement correlates with the 
 decrease of the compressive (kinetic) contribution to the stress,
 because more  room is provided to the  electronic charge to
 delocalize in between the top surface layers.
Thus the net negative change of total surface stress
 (more tensile after reconstruction)
  results from the competing
reductions of tensile and compressive contributions that are easily 
related to structural changes. In particular, while the
pre-existing tensile stress can be clearly identified as driving 
force, the change in compressive stress resulting
 from the surface rearrangement to accomodate the
surface-substrate mismatch can be interpreted as a residual
``resisting'' force of the substrate. 
The post-reconstruction stress contains
both  contributions, inextricably entangled.
It appear that the conclusion drawn from the 
post-reconstruction total stress, that the transition is not-stress driven,
 is incorrect. Further details on this 
reconstruction can be found elsewhere.\cite{me2}

The objection to this usage of the stress density 
based on its formal gauge dependence was
 considered in Sec. \ref{gauge}; there
we showed that
 perturbation-induced changes of different
 stress densities subject to the 
constraint of producing the same integrated stress
are identical to linear order in
the density, hence effectively gauge-independent.
Thus, the analysis of stress density  changes
upon reconstruction (or adsorption, etc.) 
is indeed  meaningful. 

It is worth remarking again here that 
 the analysis of stress density  changes
upon reconstruction (or adsorption, etc.) 
is  meaningful, given the proof in Sec. \ref{gauge}
that perturbation-induced changes of different
 stress densities subject to the 
constraint of producing the same integrated stress
are identical to linear order in
the density.

\section{SUMMARY}

Drawing on the theory of quantum mechanical stress,
we  introduced the stress density in density functional theory,
giving specific  prescriptions for a
practical and efficient implementation for the plane wave
ultrasoft pseudopotential method within the local-density
approximation.
In analogy with   the Chetty-Martin
energy density, the stress density provides  a spatial
 resolution of the contributions to the integrated macroscopic
 stress tensor. We applied the newly introduced concept to
a wide-ranging set of  test cases selected from surface
and interface physics. 
It is appropriate to point out that in this paper we gave just 
a flavor of the  possible applications,
 that  are far from being exhausted by those presented here. 
In perspective, the stress density is way more powerful and
predictive than the integrated macroscopic stress, 
and opens up new ways towards   an understanding
of the physics of stress-driven phenomena.

\section*{ACKNOWLEDGMENTS}
Calculations were performed mostly on the late IBM SP2
of  CRS4,  Cagliari. VF was supported by the Alexander von
Humboldt-Stiftung  during his stay at the
WSI.

\narrowtext\begin{table}
\caption{Layer-by-layer (LBL) 
decomposition of the integrated stress for an Al (111) surface slab. 
The slab is bulit up by 15 (9 atomic plus 6 vacuum) layers, labeled from 
--7 to +7. Layer 0 is the bulk-like slab center,
 and +$n$ and --$n$ layers are identical 
by  symmetry. Values are in eV/atom.
\label{fit_tab}}
\begin{tabular}{cccccccc}
   & 0 & $\pm 1$ & $\pm 2$ & $\pm 3$ & $\pm 4$ & $\pm 5$ & $\pm 6$ $\pm 7$ \\
\hline
 $T^{\rm LBL}$  & --0.779& --0.734 & --0.706 & --0.771 & --2.059& 0.729&
 0.0 \\
\end{tabular}
\end{table}

\end{multicols}

\end{document}